\documentclass[10pt,prl,twocolumn,showpacs,superscriptaddress,citeautoscript]{revtex4}
\usepackage{graphicx}
\usepackage{color}
\usepackage{amsmath}
\usepackage[normalem]{ulem}
\newcommand{\state}[2]{|#1\rangle\langle #2|}

\begin{document}
\title{Long-Lived Electronic Coherence in Dissipative Exciton-Dynamics of Light-Harvesting Complexes}
\author{Christoph Kreisbeck}
\affiliation{Institute for Theoretical Physics, University of Regensburg, 93040 Regensburg, Germany.}
\author{Tobias Kramer}
\affiliation{Institute for Theoretical Physics, University of Regensburg, 93040 Regensburg, Germany.}
\affiliation{Department of Physics, Harvard University, Cambridge, Massachusetts 02138, USA}
\begin{abstract}
The observed prevalence of oscillatory signals in the spectroscopy of biological light-harvesting complexes at ambient temperatures has led to a search for mechanisms supporting coherent transport through larger molecules in noisy environments. We demonstrate a generic mechanism supporting long-lasting electronic coherence up to $0.3$~ps at a temperature of $277$~K. The mechanism relies on two properties of the spectral density: (i) a large dissipative coupling to a continuum of higher-frequency vibrations required for efficient transport and (ii) a small slope of the spectral density at zero frequency.
\end{abstract}
\maketitle
Long-lasting oscillatory signals in optically excited light-harvesting complexes (LHCs) point towards the prevalence of a coherent electronic dynamics in molecular networks at physiological temperatures \cite{Engel2007a,Lee2007a,Collini2010a,Calhoun2009a}.
The experiments probe the electronic dynamics using two-dimensional (2d) echo-spectroscopy for a series of delay-times between the excitation pulse and the probe pulse. 
The 2d spectroscopy makes studies of the dynamics of 
dissipative systems possible and has found further applications in 
mesoscopic systems such as molecular nanotubes \cite{Abramavicius2012a} and semiconductor 
devices \cite{Kuehn2011a}.
Due to its known crystallographic structure and relative simplicity, the Fenna-Matthews-Olson (FMO) complex serves as the prototype system for studying the choreography of the energy transfer from the antenna to the reaction center of a light harvesting complex \cite{Scholes2011}.
In 2d spectra long-lasting beatings are observed, ranging from 1.2~ps at $T=150$~K to 0.3~ps at $T=277$~K \cite{Panitchayangkoon2010a}.
The interplay of coherent dynamics, which leads to a delocalization of an initial excitation arriving at the FMO network from the antenna, and the coupling to a vibronic environment with slow and fast fluctuations, has lead to studies of environmentally assisted transport in LHCs \cite{Rebentrost2009a,Plenio2008a}.

An important open question is whether coherence plays a key-role in the functioning of light-harvesting complexes \cite{Scholes2011}.
The theoretical understanding of the experiments is in its early stages and atomistic simulations based on molecular dynamics have not reached agreement \cite{Olbrich2011a, Shim2012a}.
A calculation of 2d echo-spectra based on the molecular dynamics simulation \cite{Olbrich2011b} does not show clear coherent oscillations at $T=277$~K.
One key ingredient for efficient transfer dynamics is the strong coupling to vibronic modes, which induces energy dissipation \cite{Rebentrost2009a,Plenio2008a,Kreisbeck2011a}.
For the FMO complex the thermalization occurs within picoseconds and was observed by Brixner et al.\ by the decay of diagonal-peak amplitudes to lower energies \cite{Brixner2005a}.
It has been proposed that the inclusion of the finite time scale of the reorganization 
process gives rise to long-lasting coherence in LHCs \cite{Ishizaki2009a}.
While a sluggish bath relaxation leads to prolonged population beatings in the FMO network, calculations of 2d echo-spectra show oscillations of the exciton cross-peaks for only about $1/6$th of the experimentally recorded time at $T=150$~K \cite{Hein2012a}.
For an even longer bath-relaxation cross-peak oscillations are absent already at $T=77$~K \cite{Chen2011a}.
Hence the search continues for theoretical models which support a long-lasting coherent dynamics leading to cross-peak oscillations in 2d echo-spectra and simultaneously retain the fast dissipation.
Recently, the coupling to a superposition of discrete vibronic modes rather than the existence of a purely electronic coherence has been proposed as an explanation for the detected oscillations \cite{Christensson2012a,Chin2012a}. 
However, no significant changes in the oscillations of the FMO complex have been seen in experiments which altered the vibronic modes by using native mutants, ${}^{13}$C substitution, and partial deuteration \cite{Hayes2011a}.

Here, we present an alternative mechanism which does yield long-lasting and purely electronic coherence in the presence of a strong dissipative coupling required for efficient transfer. 
The mechanism is independent of the specific exciton-system and relies on properties of the continuum part of the vibronic spectral density as opposed to discrete vibronic modes.
Realistic models of the exciton dynamics in the FMO complex have to include higher order
phonon processes as well as the finite time scale of the reorganization process \cite{Ishizaki2009a}. 
This requires to go beyond approximative 
rate equations \cite{Gardiner2004a} and non-perturbative techniques \cite{Tanimura1989a,Tanimura2006a, Nalbach2011a, Muehlbacher2008a,Koch2008a,Roden2009b,Kolli2011a}
are necessary to study the dissipative transfer dynamics.

One key parameter determining the duration of coherent oscillations is the spectral density, which encodes the the mode-dependent exciton-bath coupling.
The spectral density has important implications for the two contributions to the decoherence rate $\gamma$ which is the sum of the relaxation rate $\gamma_r$ and the pure-dephasing rate $\gamma_d$ (\cite{Weiss2008a}, ch.~21.4.2, \cite{May2004a} ch.~3.8.2),
\begin{equation}\label{eq:DephasingRate}
\gamma=\gamma_r/2+\gamma_d.
\end{equation}

Before we discuss decoherence-rates of the FMO complex, we analyze the impact of the spectral density on a two-site system with site-energies $\pm\epsilon/2$, inter-site couplings $d/2$, and $b^2=\epsilon^2+d^2$. The two rates in Eq.~(\ref{eq:DephasingRate}) are given by
$\gamma_r\approx d^2        S(b)/(2 b^2)$,
$\gamma_d=       \epsilon^2 S(0)/(2 b^2)$, 
and are linked to the spectral density $J(\omega)$ via $S(\omega)=J(\omega)\coth(\hbar\omega/(2k_B T))$.
At zero frequency, $S(0)$ is proportional to the slope of the spectral density.
We evaluate the decoherence rate for two different models of the spectral density, (i) a super-Ohmic spectral density $J_{\rm SO}\sim\omega$, and
a single-peak Drude-Lorentz spectral density $J_{\rm DL}(\omega)=2\nu\lambda\omega/(\nu^2+\omega^2)$, depicted in the Supplementary Material (SM) \cite{SM} Fig.~S\ref{fig:fig1}.

In the super-Ohmic case $S_{\rm SO}(0)=0$ and the pure-dephasing term vanishes $\gamma_d=0$ and decoherence arises only through relaxation.
The opposite regime is present for $J_{\rm DL}$ with $S_{\rm DL}(0)=4 k_B T \lambda/\nu$. 
For the parameters used in models of a sluggish bath-relaxation in the FMO complex \cite{Ishizaki2009a} 
(bath-correlation time $\nu^{-1}=50$~fs, 
reorganization energy $\lambda=40$~cm$^{-1}$, 
$\epsilon=150$~cm$^{-1}$, $d=200$~cm$^{-1}$) 
the pure-dephasing rate dominates $\gamma_{d,DL}=0.19$~cm$^{-1}$K$^{-1}$, 
over the relaxation contribution $\gamma_{r,DL}/2=0.08$~cm$^{-1}$K$^{-1}$ to the decoherence rate.

The spectral density for the FMO complex (circles in Fig.~\ref{fig:fig1}) has been extracted from fluorescence-line-narrowing experiments at $T=4$~K \cite{Wendling2000a,Renger2006a} and shows a super-Ohmic behaviour. However, previous calculations of 2d-echo spectra of seven-site FMO complex have been often performed with a single-peak Drude-Lorentz spectral density \cite{Chen2011a,Hein2012a}, and did not yield long-lasting cross-peak oscillations. 
The situation is different in the population dynamics, which has been analyzed for different spectral densities \cite{Nalbach2011a} and yields no significant differences in the duration of population beatings for different spectral density.
\begin{figure}[t]
\begin{center}
\includegraphics[width=0.82\columnwidth]{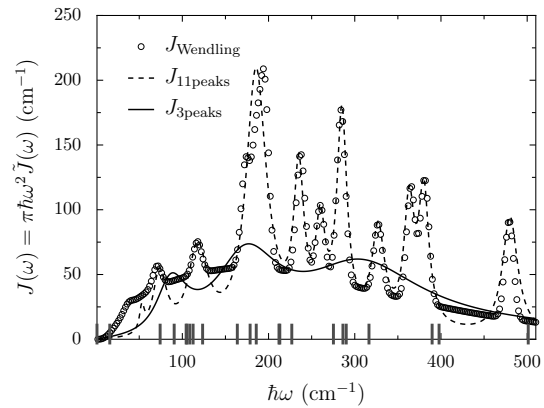}
\end{center}
\caption{\label{fig:fig1}
Spectral density of the FMO complex. Circles: measured spectral density \cite{Wendling2000a}, parametrization \cite{Renger2006a}.
Fits $J_{\rm \{3,11\} peaks}$ are used for the GPU-HEOM calculation eq.~(\ref{eq:SpecDens}). The marks on the frequency axis indicate 
the transition energies corresponding to the difference of exciton frequencies.
Parameters for $J_{\rm 3 peaks}$:
$\lambda=\{10,15,13 \}$~cm$^{-1}$, $\nu^{-1}=\{250,120,65\}$~fs, $\Omega=\{85,170,300\}$~cm$^{-1}$, $J_{\rm 11 peaks}$:
$\lambda=\{1.2, 6.4, 7.4, 15.6, 3.4, 1.8, 4, 2, 1.8, 1.9, 2\}$~cm$^{-1}$, $\nu^{-1}=\{1600, 550, 400, 370, 750, 800, 750, 600, 750, 750, 500\}$~fs, 
$\Omega=\{53, 73, 117, 185, 235, 260, 285, 327, 363, 380, 480\}$~cm$^{-1}$.
}
\end{figure}

As we show next, 2d echo-spectra are not directly derivable from population dynamics and carry along additional information about the exciton dynamics \cite{Yuen2011a}.
To demonstrate this, we calculate the exciton-dynamics for the spectral densities $J_{\rm 11 peaks}$ ($J_{\rm 3 peaks}$) shown in Fig.~\ref{fig:fig1}, which represent the experimentally determined spectral density by a superposition of $n=11\;(n=3)$ shifted Drude-Lorentz peaks \cite{Meier1999a}
\begin{equation}\label{eq:SpecDens}
J(\omega)=\sum_{k=1}^{n}\frac{\nu_k\lambda_k\omega}{\nu_k^2+(\omega\pm\Omega_k)^2}.
\end{equation}
The shifts allow us to vary the slope of the spectral density $J(\omega)$ at $\omega=0$ and to mimic the super-Ohmic behaviour seen in the experiment ($J_{\rm Wendling}$).
Compared to the single-peak Drude-Lorentz spectral-density discussed before ($\lambda=40$~cm$^{-1}$, $\nu^{-1}=50$~fs, $\Omega=0$), $S(0)$ is reduced by a factor of $13$ $(6)$. 
Both superpositions $J_{\rm \{3,11\} peaks}$  provide a good approximation of the spectral density at the transition energies corresponding to the exciton energy differences.

The Hamiltonian of the FMO complex is based on the Frenkel exciton model for LHCs \cite{May2004a} and consists of a part describing the coherent exciton dynamics
\begin{equation}\label{eq:FrenkelExc}
 \mathcal{H}_{\rm ex}=\sum_{m=1}^N \epsilon_m a_m^\dag a_m+\sum_m^N\sum_{n\ne m}^N J_{mn} a_m^\dag a_n,
\end{equation}
where $a_m^\dag$ is the creation operator of an electronic excitation of bacteriochlorophyll (BChl) $m$ with site energy $\epsilon_m$. 
The inter-site couplings between the $N$ BChls are denoted by $J_{mn}$.
The BChls are coupled to vibronic modes modeled by a set of independent harmonic oscillators $\mathcal{H}_{\rm phon}=\sum_{m,i}\hbar\omega_i b_{i,m}^\dag b_{i,m}$.
The exciton-bath coupling is given by $\mathcal{H}_{\rm ex-phon}=\sum_m a_m^\dag a_m\,u_m$ with
$u_m=\sum_i\hbar\omega_{i}d_{i}(b_{i,m}+b_{i,m}^\dag)$. 
The reorganization energy $\mathcal{H}_{\rm reorg}=\sum_m\lambda a_m^\dag a_m$ with 
$\lambda=\sum_i\hbar\omega_{i}d_{i}^2/2$ is added to the site energy $\epsilon_m$. 
For the seven-site FMO complex we take the site energies and inter-site couplings from Ref.~\cite{Renger2006a}. For simplicity we assume identical vibronic-excitonic couplings at each site and neglect correlations of vibronic modes at different sites.
The reorganization energy $\lambda\approx40$~cm${}^{-1}$ is of the same order as the excitonic energy-spacings, the bath-relaxation time, and the thermal energy $k_B T$ at $T=277$~K \cite{Ishizaki2009a,Renger2006a}.

The strong couplings of electronic and vibronic degrees require to solve the density-matrix propagation with a non-perturbative and non-Markovian approach.
The computation of the 2d echo-signal for the FMO complex necessitates about $10^3$ more time-steps than the calculation of the population dynamics and has only recently been achieved for the non-Markovian case \cite{Chen2011a,Hein2012a}.
We solve the exciton dynamics within the 
hierarchical equation of motion (HEOM)
formalism \cite{Tanimura1989a,Tanimura2006a}.
We generalize this formalism from its original single-peak Drude-Lorentz term to a superposition of shifted peaks, eq.~(\ref{eq:SpecDens}). 
The implementation on high-performance graphics processing units (GPU-HEOM) \cite{Kreisbeck2011a} allows us to advance 500 coupled hierarchical equations in lock-step and to obtain the computationally demanding 2d echo-spectra.
\begin{figure}[t]
\begin{center}
\includegraphics[width=0.82\columnwidth]{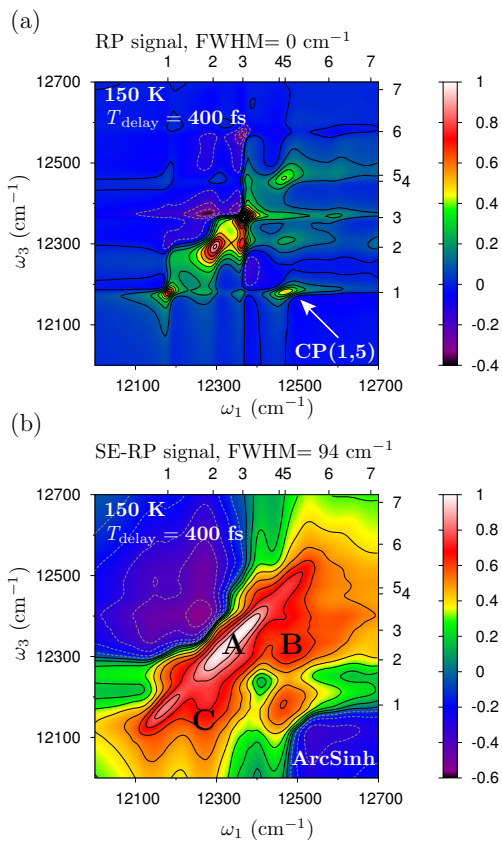}
\end{center}
\caption{\label{fig:fig2}(color online) 
Calculated 2d echo-spectra of the FMO complex at $T=150$~K for delay time $T_{\rm delay}=400$~fs, spectral-density $J_{\rm 3 peaks}$.
(a) Real part of the rephasing signal (sum of ground-state bleach, stimulated emission, and excited state-absorption) 
without static disorder (linear scale).
The grid of exciton eigenenergies is labeled $1-7$ and cross-peak CP(1,5) marked by the arrow. 
(b) Real part of the rephasing signal (stimulated emission) for static disorder FWHM$=94$~cm${}^{-1}$. 
The ArcSinh scale facilitates comparison with  \cite{Panitchayangkoon2010a}, Fig.~1.
} 
\end{figure}
We verified the convergence of GPU-HEOM by increasing the hierarchy-depth and inclusion of 
low-temperature correction terms \cite{Ishizaki2009a,Hein2012a}. 
\begin{figure}[t]
\begin{center}
\includegraphics[width=0.82\columnwidth]{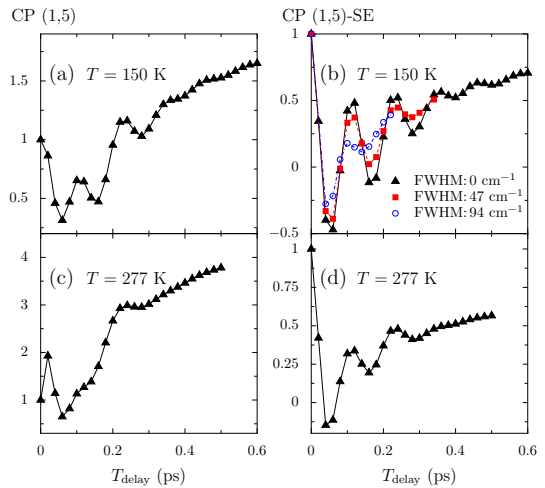}
\end{center}
\caption{\label{fig:fig3}Cross-peak~CP(1,5) oscillations of the FMO complex, real part of the rephasing signal.
The sum of ground-state bleach, stimulated emission, and excited state-absorption is shown in (a) $T=150$~K and (c) $T=277$~K. 
(b) Stimulated emission pathway at $T=150$~K for different disorder distributions FWHM$=\{0,47,94\}$~cm${}^{-1}$.
(d) Stimulated emission pathway at $T=277$~K (no disorder).
} 
\end{figure}

A typical 2d FMO-spectrum at $T=150$~K and delay time $T_{\rm delay}=400$~fs for the spectral density $J_{\rm 3peak}$ is shown in Fig.~\ref{fig:fig2}(a).
The 2d echo-spectra are calculated as detailed in Ref.~\cite{Hein2012a} by evaluating the third-order response function, including excited-state absorption \cite{Mukamel1999a} for the non-disorder averaged results.
We mimic rotational averaging by probing along eight molecular orientations with respect to the laser-field polarization.

The inclusion of static disorder leads to a broadening of the peaks as shown in Fig.~\ref{fig:fig2}(b) for the stimulated-emission rephasing pathway and a Gaussian disorder-ensemble of 200 realizations with FWHM$=94$~cm${}^{-1}$ added to the diagonal site-energies. The disordered peaks in Fig. 2(b) (which does not include the excited-state absorption) are somewhat narrower than in the experimental data \cite{Panitchayangkoon2010a}, Fig.~1C. 
In Fig.~\ref{fig:fig2}, we reference the peaks on the diagonal (DP) and the cross-peaks (CP) by their location on the 
exciton eigenenergy-grid.
Disorder leads to a broad central region where the peaks coalesce (compare Fig.~\ref{fig:fig2}(a) and (b)), and amplitude is present in regions A, B, and C, in reasonable agreement the experimental data in \cite{Panitchayangkoon2010a}, Fig.~1C.
The calculated 2d-spectra differ significantly from the single-peak Drude-Lorentz results shown in the SM \cite{SM}, Fig.~S5 and in \cite{Hein2012a}.
Besides the broader line-shapes in the single-peak Drude-Lorentz case, the dynamics of the diagonal-peak amplitudes is altered and an upward amplitude transfer from DP(2,2) to DP(3,3) is visible for $J_{\rm 3 peaks}$, which is in line with the experimental data at $T=77$~K \cite{Brixner2005a}, but does not exist in the single-peak Drude-Lorentz case.

\begin{figure}[t]
\begin{center}
\includegraphics[width=0.82\columnwidth]{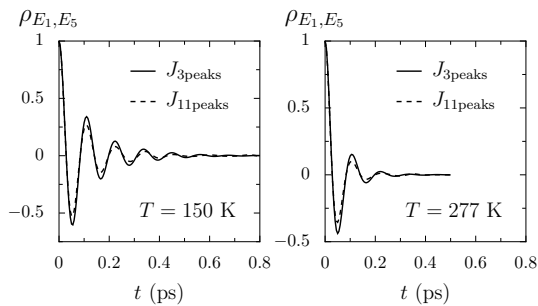}
\end{center}
\caption{\label{fig:fig4}
Exciton dynamics for the FMO with the spectral densities $J_{\rm 3 peaks}$ and $J_{\rm 11 peaks}$. 
Time evolution of the coherence $\rho_{E_1,E_5}(t)=\langle E_1|\rho(t)| E_5\rangle$.
The duration of coherent oscillations of cross-peak~CP(1,5) shown in Fig.~\ref{fig:fig3} is in good agreement with the coherence life times $\rho_{E_1,E_5}(t)$.
}
\end{figure}

To determine the coherence-time of cross-peak oscillations, we analyze the changes in the amplitude of the cross-peak~CP(1,5) marked with the arrow in Fig.~\ref{fig:fig2}(a).
The oscillatory signal of CP(1,5) for $J_{\rm 3peaks}$ shows modulations for the rephasing pathways (Fig.~\ref{fig:fig3}(a)). 
The coherent nature of the oscillations is best analyzed in the stimulated-emission (SE) pathway in Fig.~\ref{fig:fig3}(b) which shows oscillations at $T=150$~K ($277$~K) for delay-times $T_{\rm delay}$ up to $0.6$~ps ($0.3$~ps). 
The excited-state absorption included in Fig.~\ref{fig:fig3}(a) neglects correlation between excitons and almost compensates the stimulated-emission, resulting in a larger non-oscillatory background. 
The addition of static disorder diminishes the oscillations to some extent, but the oscillatory signal prevails as shown for the SE-pathways in Fig.~\ref{fig:fig3}(b) at $T=150$~K.
In the single-peak Drude-Lorentz case discussed in \cite{Hein2012a} and the SM \cite{SM}, Fig.~S4, the oscillations decay two times faster and no long-lasting oscillations remain in the stimulated-emission signal at  $T=277$~K.

The information of the exciton lifetime is already incorporated in the simpler to calculate single exciton dynamics. 
To this end, we prepare the system in an initial density matrix $\rho(t=0)=\state{E_i}{E_j}$ given in terms of two eigenstates 
$\{|E_i\rangle, |E_j\rangle\}$ of ${\cal H}_{\rm ex}$. 
In the 2d echo-spectra setup such a delocalized exciton state is reached after the first two laser pules illuminate the sample. 
The life time of the oscillatory components of cross-peak CP(1,5) of the 2d echo-spectra in Fig.~\ref{fig:fig3} is in good agreement with the life time of the corresponding coherences  
$\rho_{E_1,E_5}(t)=\langle E_1|\rho(t)| E_5\rangle$ shown in Fig.~\ref{fig:fig4}.

To isolate the impact of the vibronic mode-distribution we compare the coherences for the spectral densities $J_{\rm 3 peaks}$ and $J_{\rm 11 peaks}$ in Fig.~\ref{fig:fig4}.
The coherences are almost unchanged by switching from $J_{\rm 3peaks}$ to $J_{\rm 11peaks}$. 
Specific modes in the peaked spectral density as the one near $180$~cm$^{-1}$ induce beatings in the coherences with small amplitude $\le0.01$.
In contrast to $\delta$-shaped vibronic modes, the peaks in $J_{\rm 11peaks}$ which mimic the experimental data, have only a minor influence to the coherences.
However, this does not imply that higher-frequency modes are unimportant. 
The relaxation to thermal equilibrium, which is prominently visible in 2d-echo spectra, relies on a large enough coupling of the exciton-system to the vibronic continuum.
The spectral densities studied have a reorganization energy ${\lambda}\geq35$~cm$^{-1}$ which lead to a thermalization after $5$~ps at $T=77$~K as observed for the FMO complex \cite{Brixner2005a}. 

For efficient energy-transfer it is important to verify that the decoherence rate $\gamma$ still supports a thermalization of the exciton system within a few picoseconds. 
This is indeed the case for $J_{\rm 3 peaks}$ and $J_{\rm 11 peaks}$ and shows that despite the reduced slope of the spectral density, the system is still strongly coupled to the vibronic environment.
Compared to the single-peak Drude-Lorentz form, the spectral densities $J_{\rm \{3,11\} peaks}$ support even a faster transfer from the initial BChl~6 to the target BChl~3.
In agreement with the experimental observations, the 2d spectra calculated here for the spectral density $J_{\rm 3 peaks}$ do support long-lasting oscillations of the \textit{cross-peaks}.

In conclusion, we have shown that two seemingly contradictory requirements, namely (i) strong coupling needed for a fast thermalization and (ii) the observation of long-lasting coherent oscillations are two sides of the same coin.
The careful modeling of the continuous part of the spectral density towards zero frequency is essential for the prevalence of coherence in molecular networks since it determines the pure-dephasing rate $\gamma_d$.
By combining the exciton Hamiltonian with a spectral density with the small slope at $\omega=0$ seen in experiments, 
we find a much better agreement of the theoretically calculated 2d spectra and the experimental data once static disorder is included.
The vanishing pure-dephasing rate of a super-Ohmic spectral density implies that the decoherence is driven by the relaxation rate $\gamma_r$.
This is exactly the opposite behaviour compared to models based on the single-peak Drude-Lorentz spectral density \cite{Ishizaki2009a}.
Furthermore, we find only a very small influence of the additional peaks in the spectral density $J_{\rm 11 peaks}$ on the coherences.
The cross-peak oscillations require a calculation of 2d-echo spectra beyond the weak-coupling expansion for structured spectral densities, which becomes feasible by the efficient GPU-HEOM algorithm.
\acknowledgments
We thank B.~Hein for help with calculating 2d-spectra and M.~Rodr\'iguez for helpful discussions. This work is supported by the DFG within the Emmy-Noether program (KR~2889). 
Time on the Harvard SEAS ``Resonance'' GPU cluster and support by the SEAS Academic Computing team are gratefully acknowledged.

\appendix
\onecolumngrid

\section{Appendix}

\section{Comparison of exciton dynamics for different spectral densities}

\begin{figure}[b]
\begin{center}
\includegraphics[height=0.325\columnwidth]{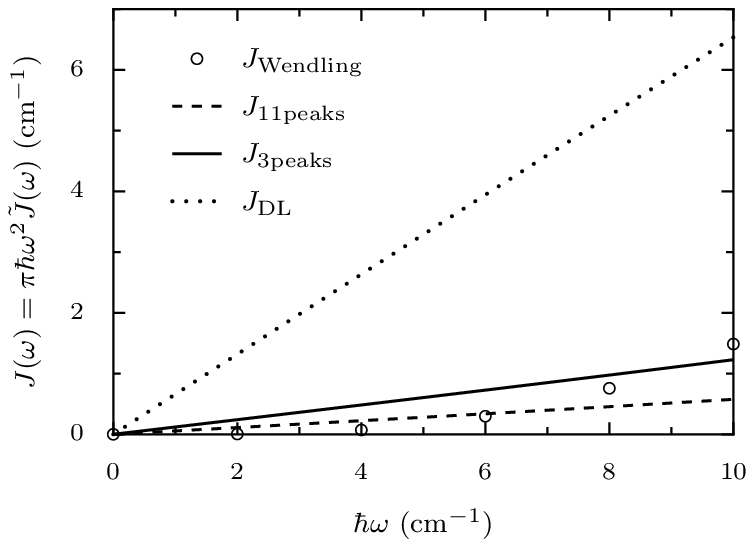}\hfill
\includegraphics[height=0.325\columnwidth]{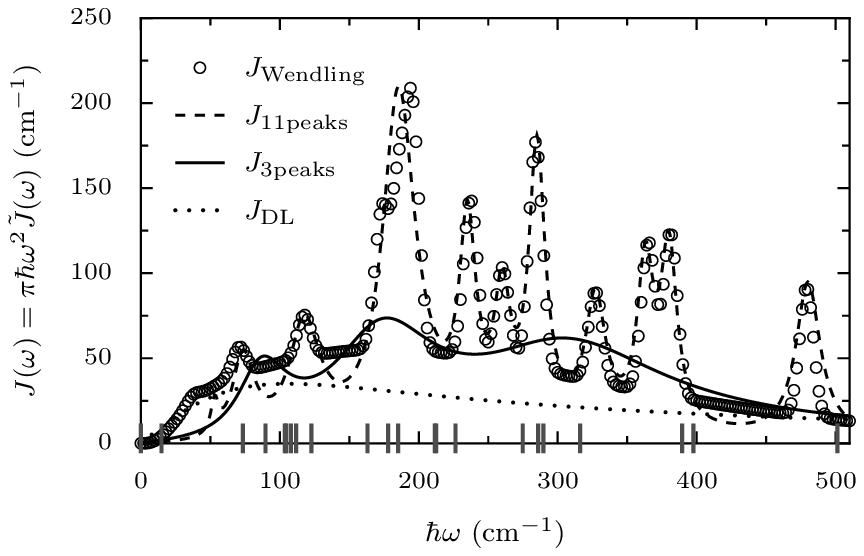}
\end{center}
\caption{\label{fig:sfig1}
Left panel: low frequency onset of the spectral density of the FMO complex. Right panel: global view.
Circles: measured spectral density \cite{Wendling2000a}, parametrization \cite{Renger2006a}.
Solid line $J_{\rm 3 peaks}$, dashed line $J_{\rm 11 peaks}$, dotted line $J_{\rm DL}$, parameters given in the main text.
}
\end{figure}

The exciton dynamics is strongly affected by the spectral density and the dephasing rate in Eq.~(1) depends on the slope of the spectral density towards zero frequency.
Fig.~\ref{fig:sfig1} shows a close up of spectral densities discussed in the main text, and in additions includes for convenient reference the single-peak Drude-Lorentz density 
(dotted line, $J_{\rm DL}$), used in some previous studies \cite{Ishizaki2009a,Chen2011a,Hein2012a}.
\pagebreak

\subsection{Population dynamics and coherences}

\begin{figure}[t]
\begin{center}
\includegraphics[width=0.6\columnwidth]{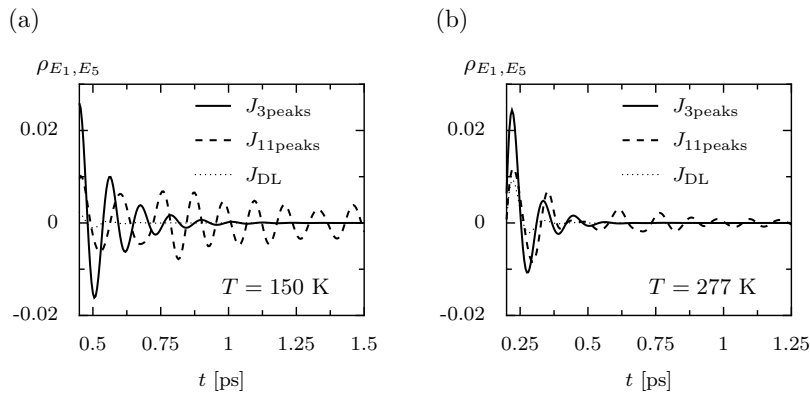}
\end{center}
\caption{\label{fig:sfig4zoom}
Detail of the long-time evolution of the coherence $\rho_{E_1,E_5}(t)=\langle E_1|\rho(t)| E_5\rangle$, which shows small-amplitude oscillations for $J_{\rm 11peaks}$ related to the peaks of the spectral density $J_{\rm 11peaks}$ located at higher frequencies. For $J_{\rm DL}$ (dotted line), almost no coherences prevail.
}
\end{figure}

\begin{figure}[b]
\begin{center}
\includegraphics[width=0.485\columnwidth]{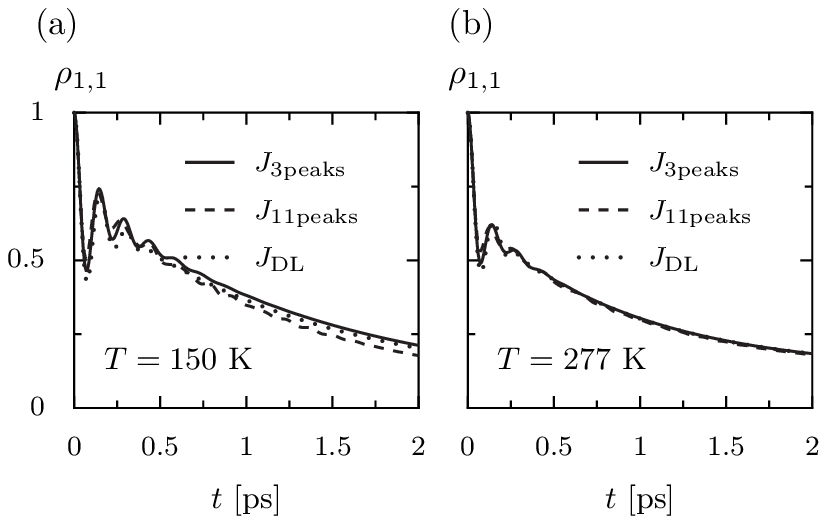}\hfill
\includegraphics[width=0.485\columnwidth]{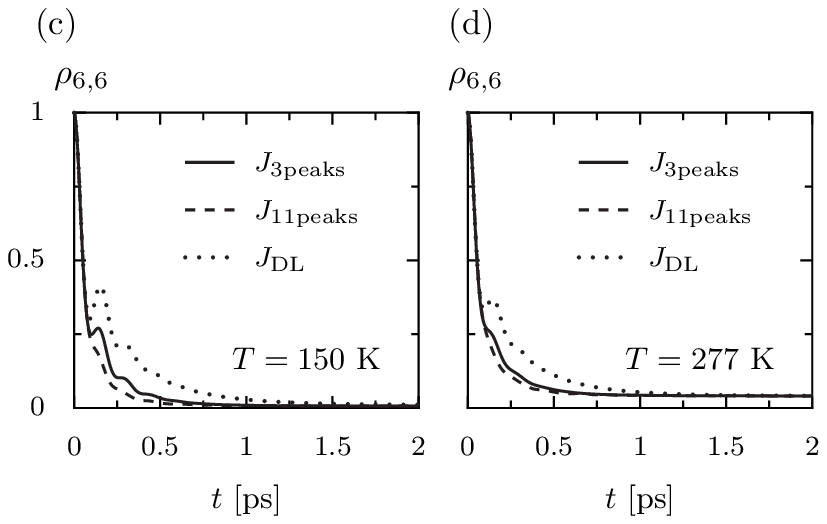}
\end{center}
\caption{\label{fig:sfig2}
Population dynamics for an initial population at sites $1$ and $6$ for the $3$ spectral densities shown in Fig.~\ref{fig:sfig1} and for temperatures $T=150$~K and $T=277$~K.
The relaxation proceeds on very-similar time-scales for the two spectral densities $J_{\rm 3 peaks}$ and $J_{\rm 11 peaks}$.
}
\end{figure}

The additional peaks present in $J_{\rm 11peaks}$ at higher frequencies, which are not contained in $J_{\rm 3peaks}$, affect the coherences of the off-diagonal elements $\rho_{E_1,E_5}$ only weakly, Fig.~\ref{fig:sfig4zoom}. 
The influence of the additional peaks in the spectral density $J_{\rm 11peaks}$ becomes visible as small-amplitude oscillations for longer delay times (Fig.~\ref{fig:sfig4zoom}).
This shows how the initially electronic coherence gets transformed into oscillations related to the details of the sharp peaks in the spectral density $J_{\rm 11peaks}$.
This finding is in-line with the theoretical calculations for spectral-densities with added $\delta$-peaked vibronic modes \cite{Christensson2012a}. 
The same observation applies to the relaxation process, which is shown in Fig.~\ref{fig:sfig2} for a population dynamics starting either at site $1$ or at site $6$. 
For comparison, the results for the Drude-Lorentz spectral density (dotted curve) is also included, which deviates most significantly for $\rho_{66}$.

\pagebreak

\subsection{Differences in 2d spectra}

\begin{figure}[t]
\begin{center}
\includegraphics[width=0.85\columnwidth]{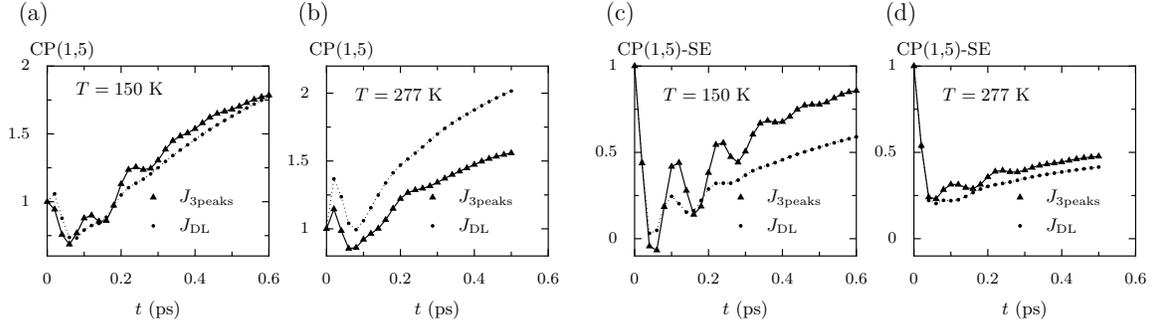}
\end{center}
\caption{\label{fig:sfig3}
Exciton dynamics for the FMO complex for two different spectral densities $J_{\rm 3 peaks}$ and $J_{\rm DL}$. (a) Real part of cross-peak~CP(1,5) of the signal (rephasing and non-rephasing) at $T=150$~K and
$T=277$~K. 
(b)  Coherent oscillations of the stimulated emission (SE) part of cross-peak~CP(1,5).
} 
\end{figure}

The cross-peak oscillations depend strongly on the initial slope of the spectral densities towards zero frequency. This is reflected in Fig.~\ref{fig:sfig3}, which compares the cross-peak oscillations for the two different spectral densities $J_{\rm 3 peaks}$ and $J_{\rm DL}$, for which the dephasing- and relaxation contributions to the decoherence rate are reversed in magnitude.
Fig.~\ref{fig:sfig3}(b) shows that for $J_{\rm DL}$ (dotted lines) the cross-peak oscillations are largely reduced at $T=150$~K and almost completely absent at $T=277$~K, while the super-Ohmic onset of $J_{\rm 3 peaks}$ still leads to visible oscillations at $T=277$~K (solid lines).
\enlargethispage{4.5cm}
Moreover, also the amplitudes of the diagonal-peaks show pronounced differences for the two spectral densities. The spectral density $J_{\rm 3 peaks}$ transfers the highest diagonal-peak amplitude upwards  with increasing delay time from DP(2,2) to DP(3,3) as illustrated in the left panels of Fig.~\ref{fig:sfig3}. This shift is also seen in the experimental data at $T=77$~K \cite{Brixner2005a}, Figure 1, panels a,b.\\
In the Drude-Lorentz case (Fig.~\ref{fig:sfig3}, right panels) the highest amplitude is located at diagonal-peak DP(2,2) for all delay-times, which corresponds to previous models (see Fig.~2 in \cite{Bruggemann2007}), but differs from the experimental data and the result for $J_{\rm 3 peaks}$.
\begin{figure}[b]
\begin{center}
\includegraphics[width=0.65\columnwidth]{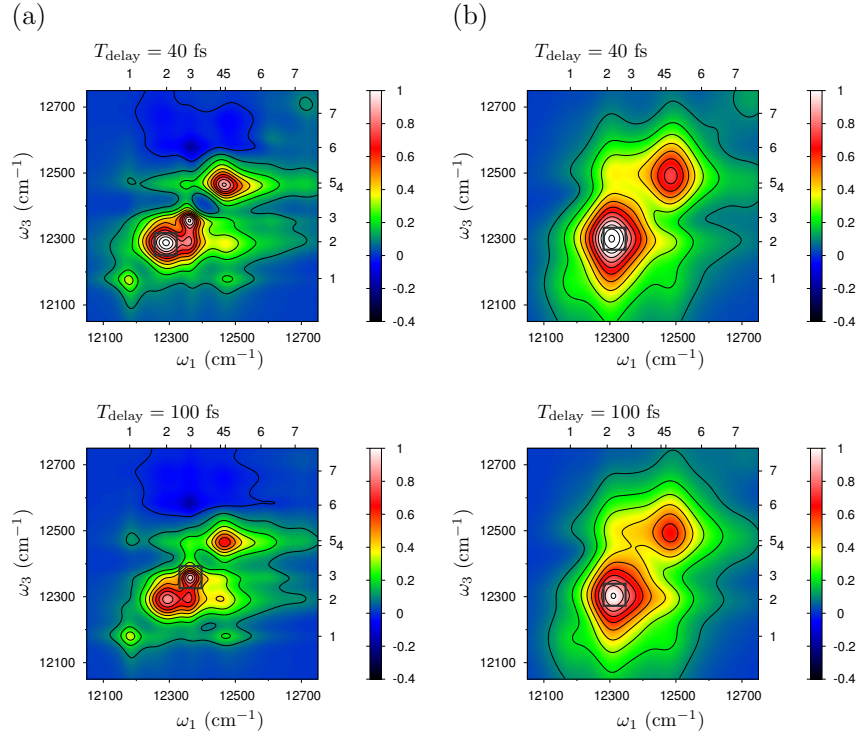}
\end{center}
\caption{\label{fig:sfig4}(color online) Calculated 2d echo-spectrum of the FMO complex at $T=277$~K for the delay times $T_{\rm delay}=\{40,100\}$~fs.
Shown is the real part of the rephasing and non-rephasing signal. The square marks the diagonal peak (DP) with the highest amplitude. 
Left panels: spectral density $J_{\rm 11 peaks}$, showing amplitude shifts from DP(2,2) at $T_{\rm delay}=40$~fs to DP(3,3) for $T_{\rm delay}=100$~fs, right panels: spectral density $J_{\rm DL}$, highest amplitude remains at DP(2,2).
} 
\end{figure}



\end{document}